\documentclass[english,aps,PRE,twocolumn]{revtex4}
\usepackage[T1]{fontenc}
\usepackage[latin9]{inputenc}
\usepackage{amssymb}
\usepackage{graphicx}
\usepackage{esint}

\makeatletter
\@ifundefined{textcolor}{}
{%
 \definecolor{BLACK}{gray}{0}
 \definecolor{WHITE}{gray}{1}
 \definecolor{RED}{rgb}{1,0,0}
 \definecolor{GREEN}{rgb}{0,1,0}
 \definecolor{BLUE}{rgb}{0,0,1}
 \definecolor{CYAN}{cmyk}{1,0,0,0}
 \definecolor{MAGENTA}{cmyk}{0,1,0,0}
 \definecolor{YELLOW}{cmyk}{0,0,1,0}
}

\makeatother

\usepackage{babel}
\begin{document}

\title{General formulation of Luria-Delbrück distribution of the number
of mutants.}

\author{Bahram Houchmandzadeh}

\affiliation{CNRS, LIPHY, F-38000 Grenoble, France\\
Univ. Grenoble Alpes, LIPHY, F-38000 Grenoble, France}
\begin{abstract}
The Luria-Delbrück experiment is a cornerstone of evolutionary theory,
demonstrating the randomness of mutations before selection. The distribution
of the number of mutants in this experiment has been the subject of
intense investigation during the last 70 years. Despite this considerable
effort, most of the results have been obtained under the assumption
of constant growth rate, which is far from the experimental condition.
We derive here the properties of this distribution for arbitrary growth
function, for both the deterministic and stochastic growth of the
mutants. The derivation we propose uses the number of wild type bacteria
as the independent variable instead of time. The derivation is surprisingly
simple and versatile, allowing many generalizations to be taken easily
into account.
\end{abstract}
\maketitle

\section{Introduction.}

The Darwin-Wallace theory of evolution rests upon mutation of living
organisms and their selection. In their seminal article \cite{Luria1943},
Luria and Delbrück (LD) described an experiment demonstrating the
randomness of mutational events \emph{before} the selection process.
The experiment consists of growing $C$ cultures of bacteria in parallel
in identical environments, beginning with a small number $N_{0}$
(typically $10^{3}$) in each batch. After a sufficient growth period,
the cultures saturate and the number of wild type bacteria reaches
$N$ (typically $10^{9}-10^{10}$). Each culture is then tested against
an antibacterial agent, a phage virus in the LD case, and the number
of surviving bacteria arising from mutation in the cultures is counted
by a plating method. If $C$ is large, the probability $P(m)$ of
having $m$ mutants can be experimentally determined; in practice,
$C$ cannot be large and therefore only statistical quantities such
as the mean and the variance of the number of mutants can be estimated.

The LD experiment has spurred a large interest and many authors have
developed increasingly refined models to estimate statistical properties
of the random variable $X$ of the number of mutants, such as its
cumulant/probability/moment generating functions (cgf, pgf, mgf),
from which the mutation rate or probability can be estimated. The
pioneering authors were Lea and Coulson\cite{Lea1949}, Armitage \cite{Armitage1952},
Bartlett\cite{Bartlett1955}, Crump and Hoel\cite{CRUMP1974}, Mandelbrot\cite{Mandelbrot1974},
Sarkar, Ma and Sandri\cite{Sarkar1992} who set the LD distribution
on a solid mathematical ground, generalized the model to take into
account stochastic growth of the mutant and the wild type, and developed
algorithms to estimate the mutation rate. The works of these and other
authors have been reviewed in an elegant article by Zheng\cite{Zheng1999}
which also contains original results and corrections of some of the
errors contained in the previous works. These investigations have
been extended during the last 15 years by authors such as Angerer\cite{Angerer2001a},
Dewanji et al\cite{Dewanji2005}, Ycart\cite{Ycart2013}, Kessler
and Levine\cite{Kessler2013,Kessler2014}. A description of some of
these more recent works will be given in the following sections.

The fundamental LD experiment is now currently used to estimate mutation
rates in various setups such as antibiotic resistance or experimental
investigation of the evolutionary process\cite{Rosche2000,Sniegowski1997a,pal2007}. 

However, nearly all of the existing computations have focused on the
exponential growth (either deterministic or stochastic) of the wild
type and mutant bacteria, although Dewanji et al.\cite{Dewanji2005}
have extended these results to Gompertz growth. The reason behind
this choice is that in these formulations, an explicit expression
for the number of wild type (WT) cells as a function of time is needed
in order to compute the statistical properties of the number of mutants.

The assumption of constant growth rate is however too restrictive.
Experimentally, the growth is never exponential but follows a Monod
curve\cite{Monod1949} : the growth rate is not constant, but begins
with a value close to zero (called the lag phase), increases gradually
to a maximum value and then decreases as the number of bacteria increases,
to finally reach zero when the culture is saturated. Various functions
(logistic, Gompertz, Richards, Stannard, ...) are used in the literature
to model the growth curve and their relevance has been studied in
depth by Zwieteting et al. \cite{Zwietering1990}. 

The \emph{real time} however is not the relevant independent variable
in terms of which the system may be described. The WT population grows
from an initial number of cells $N_{0}$ to reach a final value $N$.
Each time a WT cell divides, there is a small probability $\nu$ that
a mutant having the desired trait (phage or antibiotic resistance)
appears. It does not matter how much time the system spends between
WT population size $n$ and $n+1$, but only the fact that once a
division has taken place, a mutation may appear. For the mutants growing
in the same environment as the WT, their growth curve will be similar
(but not necessarily equal) to that of the WT. The only quantities
that are indeed measured in an experiment are the number of mutants
$m$, the initial WT population $N_{0}$ and the final population
$N+m$. Even though the growth curve can theoretically be measured,
its determination is cumbersome and, as we will see, not relevant. 

In this paper, we shall use the WT population size $n$ as the independent
variable. It appears that this formulation of the LD distribution
is surprisingly simple and applies to any growth curve for the bacteria,
including of course exponential growth. As in the case of classical
derivation of the LD distribution, this derivation is valid in the
limit of large final population size, which is typical of experiments
involving bacteria. The formulation is versatile and can take into
account many generalizations, some of which are considered in this
article. A similar approach has been taken by Kessler and Levine\cite{Kessler2013,Kessler2014}
when solving directly for the Master equation governing the dynamics
of the mutant population. 

This article is organized as follows: in the next section, we present
the basic concept for the simple case of deterministic growth of both
WT and mutant bacteria, where only mutations apparitions are random.
We present some generalizations, such as different growth rates for
WT and mutants and non-constant mutation probability. In the following
section, we generalize the model to the case of stochastic growth
of the mutant, where we consider (i) a linear birth process for the
mutant and (ii) a random relative growth rate for the mutant. We stress
that in the following computations, the growth rate is not constant,
but can have an arbitrary form. The last section is devoted to a discussion
of possible extensions of this work and conclusions. An appendix,
containing straightforward mathematical derivations is included in
order to make this article self sufficient.

\section{Deterministic model.}

\subsection{Equal growth of WT and mutant.\label{sub:Equal-growth-of}}

Consider a culture of WT bacteria growing from size $N_{0}$ to size
$N$. The growth curve can be as general as possible assuming that
no death event takes place. Let $X_{n}$ be the random variable describing
the occurrence of a mutant when the WT population increases from $n$
to $n+1$. Throughout this paper, we use the term mutant to designate
an individual which acquire a trait (e.g. phage or antibody resistance)
which will be tested once the growth is stopped. 

Denoting the mutation \emph{probability} by $\nu$, 
\begin{equation}
\mbox{Pr}(X_{n}=0)=1-\nu\,\,;\,\,\mbox{Pr}(X_{n}=1)=\nu\label{eq:XnProba}
\end{equation}

This may seem to be an approximate description, because if during
a cell division, a mutation has occurred, obviously the number of
WT cells has not increased from $n$ to $n+1$. A more precise formulation
would be $\mbox{Pr}(X_{n}=k)=(1-\nu)\nu^{k}$, \emph{i.e.,} the number
of mutants when the WT population increases by one unit is geometrically
distributed. However, as $\nu\ll1$ ($\nu$ is usually of the order
of $10^{-8}$), we will use the relation (\ref{eq:XnProba}) to describe
the random variable $X_{n}$. The generalization to the geometric
distribution of $X_{n}$ is straightforward (see appendix  \ref{sub:Geometric-distribution-of}).
Note that most formulations of LD distribution use the above approximation. 

\begin{figure}
\begin{centering}
\includegraphics[width=0.85\columnwidth]{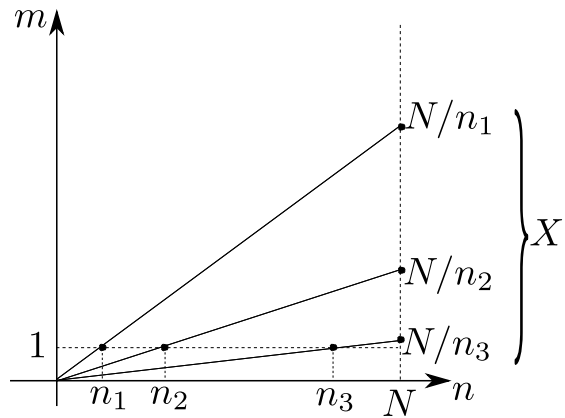}
\par\end{centering}

\protect\caption{The number of mutants $X$ at WT population size $N$ is the sum of
the contributions of mutants lineages appearing at WT population size
$n_{i}$. The size of the lineage of mutant $i$ appearing at WT population
size $n_{i}$ is $m_{i}(n)=n/n_{i}$ for $n\ge n_{i}$. The $i-$th
mutant, appearing at size $1$ at WT population $n_{i}$ will be present
at size $N/n_{i}$ in the final population. \label{fig:fundconcept}}

\end{figure}

We assume in this subsection that both mutant and the WT population
follow a deterministic, equal growth. We do not assume the growth
rate to be a constant. As the mutants are similar to the WTs, a mutant
appearing in one copy at WT population size $n$ will contribute $N/n$
to the number of mutants when the WT population reaches size $N$
(figure \ref{fig:fundconcept}). In other words, the proportion of
the number of this mutant to the number of WT population will remain
constant.

Let $Y_{n}^{N}=(N/n)X_{n}$ be the contribution of this mutant to
the final number of mutants $X$, when the WT population reaches size
$N$. Then
\[
X=\sum_{n=N_{0}}^{N}Y_{n}^{N}
\]
As the mutant occurrences are independent random variables, the moment
(mgf) and cumulant generating functions (cgf) are 
\begin{eqnarray*}
\phi(s) & = & \left\langle e^{sX}\right\rangle =\prod_{n=N_{0}}^{N}\left\langle e^{sY_{n}^{N}}\right\rangle \\
\psi(s) & = & \log\left(\left\langle e^{sX}\right\rangle \right)=\sum_{n=N_{0}}^{N}\log\left(\left\langle e^{sY_{n}^{N}}\right\rangle \right)
\end{eqnarray*}
On the other hand, by its very definition, 
\[
\left\langle e^{sY_{n}^{N}}\right\rangle =1-\nu+\nu e^{sN/n}
\]
which gives the cgf as 
\begin{eqnarray}
\psi(s) & = & \sum_{n=N_{0}}^{N}\log\left(1-\nu+\nu e^{sN/n}\right)\nonumber \\
 & = & N\int_{x_{0}}^{1}\log\left(1-\nu+\nu e^{s/x}\right)dx\label{eq:cgf-deterministic}
\end{eqnarray}
where in the second expression, we have used the continuous approximation
for the sum, $x=n/N$ and $x_{0}=N_{0}/N$. The relative error in
using the continuous approximation is at most ${\cal O}\left(1/N_{0}\log(x_{0})\right)$.

Note that this derivation is analogous to the filtered Poisson process
derivation used when the problem is formulated in real time (\cite{Zheng1999},eq.
14). However, because the problem is formulated in terms of WT population
size, the propagator is simply a straight line regardless of the growth
rate function (figure \ref{fig:fundconcept}). 

Expanding the expression (\ref{eq:cgf-deterministic}) to the first
order in $\nu$ and restricting the domain of definition to $s\lesssim-x_{0}\log\nu$,
\begin{equation}
\psi(s)=-\theta\phi+\theta\int_{x_{0}}^{1}e^{s/x}dx+O(\nu^{2})\label{eq:psiapprox}
\end{equation}
Where $\theta=N\nu$ and $\phi=1-x_{0}$. For the particular case
of exponential growth, the expression (\ref{eq:psiapprox}) for the
cgf has been obtained by Crump and Hoel \cite{CRUMP1974} and in closed
form by Zheng (\cite{Zheng1999},eq.14). Note that for Zheng, $N=exp(\beta t)$,
$i.e.$ the initial number of WT bacteria is $1$. 

The first two cumulant coefficients $\kappa_{p}=\psi^{(p)}(0)$ are
then
\begin{eqnarray}
\kappa_{1} & = & -\theta\log x_{0}\label{eq:kappa1}\\
\kappa_{2} & = & \frac{\theta\phi}{x_{0}}\label{eq:kappa2}
\end{eqnarray}
These expressions for the average ($\kappa_{1}$ ) and the variance
($\kappa_{2}$ ) have been obtained originally by LD for the exponential
growth of bacteria \cite{Luria1943}. As we see here, the hypothesis
of exponential growth is superfluous and the expressions (\ref{eq:kappa1}-\ref{eq:kappa2})
are valid for arbitrary growth curves. To the leading order in $\nu$
and $x_{0}$ , the general expression for cumulant coefficients is
\begin{equation}
\kappa_{p+1}=\frac{\theta}{px_{0}^{p}}\,\,\,\,\,p>0\label{eq:kappap}
\end{equation}
 These expressions are known for the special case of exponential growth
(\cite{Zheng1999},eq.9 for equal growth rate).

\subsection{Different growth of WT and mutant.\label{sub:Different-growth}}

Let us now consider the case where WT ($n$) and mutants ($m$) (once
they have appeared) have similar but different growth rates:
\[
\frac{dn}{dt}=\alpha(n,t)n\,\,;\,\,\frac{dm}{dt}=c\alpha(n,t)m
\]
where $c$ is a constant. We do not specify any particular form for
the growth rate, but we suppose that the mutant follows the same law
as the WT, within a constant multiplicative factor. This is the case
for example where the resources are depleted by the growth of the
bacteria, and the mutant is inferior to the WT for its duplication. 

\begin{figure}
\begin{centering}
\includegraphics[width=0.85\columnwidth]{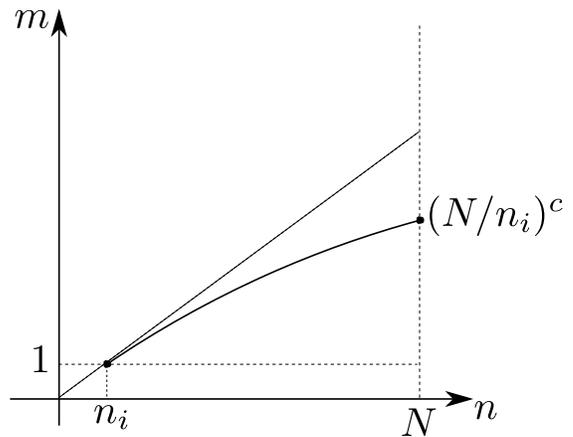}
\par\end{centering}

\protect\caption{Contribution of mutants with different growth rate. The size of the
lineage of mutant $i$ appearing at WT population size $n_{i}$ is
$m_{i}(n)=(n/n_{i})^{c}$ for $n\ge n_{i}$. \label{fig:difgrowthrate}}
\end{figure}

Time can be eliminated between the above equations: $dm/dn=c(m/n)$.
A mutant appearing at one copy when the population size is $n$ will
contribute $(N/n)^{c}$ to the final number of mutants (figure \ref{fig:difgrowthrate}).
 The computation of the preceding section can be repeated and leads
to 
\[
\psi(s)=N\int_{x_{0}}^{1}\log\left(1-\nu+\nu e^{s/x^{c}}\right)dx
\]
Keeping only the leading term in $\nu$ and $x_{0}$, for $c\ne1/p$,
the $p-$th cumulant coefficient is given by
\[
\kappa_{p}=\frac{\theta}{cp-1}(x_{0}^{1-cp}-1)
\]
The case $c=1/p$ can be recovered from the above formula by taking
the limiting value for $c\rightarrow1/p$ and reads
\[
\kappa_{p}=-\theta\log x_{0}
\]
For the particular case of exponential growth ($\alpha(n,t)=\alpha$
), this is the expression given by Zheng (\cite{Zheng1999},eq.9).

\subsection{Variable mutation probability.\label{sub:Variable-mutation-probability.}}

In most models the mutation probability $\nu$ is considered to be
a constant and independent of time, \emph{i.e., }population size.
This is a sound hypothesis when the mutation involves only point-mutations
on the chromosome. However, traits such as virus or antibiotic resistance
may involve many point mutations before the trait is functional. Bacteria
at the end of the growth process, having achieved more divisions,
may be more prone to mutate to the given trait than bacteria at the
early stage of the growth. A crude approximation of the above phenomena
will be a mutation rate that depends on the population size $\nu=\nu(n)$.
The formulation for the number of mutants of the previous section
does not suppose a constant rate of mutation and the relation \ref{eq:cgf-deterministic}
for $\psi(s)$ remains valid. The first two cumulants are given by
\begin{eqnarray*}
\kappa_{1} & = & N\int_{x_{0}}^{1}\frac{\nu(x)}{x}dx\\
\kappa_{2} & = & N\int_{x_{0}}^{1}\frac{\nu(x)-\nu^{2}(x)}{x^{2}}dx
\end{eqnarray*}

\section{Stochastic growth of mutants.}

\subsection{General discussion.\label{sub:General-discussion.}}

Until now, we have considered the deterministic propagation of the
mutant from its appearance at population size $n$ to its final value
at population size $N$. We will denote the propagator as $K_{n}^{N}$.
The contribution of the mutant appearing at WT population size $n$
to the final population $N$ was expressed as 
\[
Y_{n}^{N}=X_{n}K_{n}^{N}
\]
For the deterministic case, $K_{n}^{N}=N/n$ and the mgf of $Y_{n}^{N}$
was simply 
\[
\left\langle e^{sY_{n}^{N}}\right\rangle =1-\nu+\nu e^{sN/n}
\]

\begin{figure}
\begin{centering}
\includegraphics[width=0.85\columnwidth]{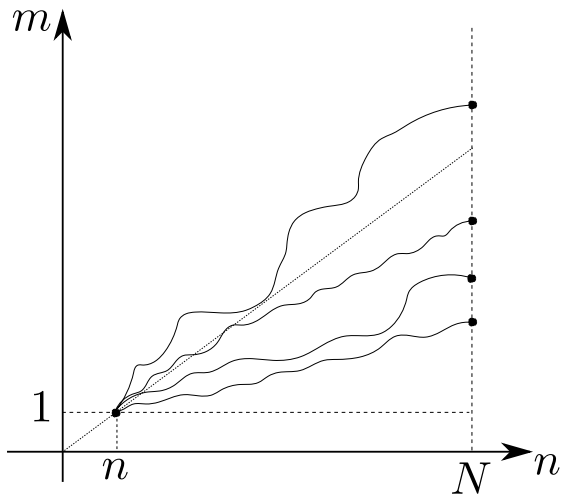}
\par\end{centering}

\protect\caption{The stochastic propagator $K_{n}^{N}$ of the mutant appearing at
WT population size $n$.\label{fig:Stochastic-propagator}}
\end{figure}

We will now consider for the mutant a stochastic propagator $K_{n}^{N}$
(figure \ref{fig:Stochastic-propagator}). Because $X_{n}$ takes
only the values 0 or 1, 

\begin{equation}
\left\langle e^{sY_{n}^{N}}\right\rangle =\left\langle e^{sX_{n}K_{n}^{N}}\right\rangle =1-\nu+\nu\left\langle e^{sK_{n}^{N}}\right\rangle \label{eq:charYn}
\end{equation}
(see appendix \ref{sub:appa}). Therefore, all the discussion of the
preceding section naturally generalizes to stochastic propagation
and the cgf for the number of mutants at population size $N$ is 
\begin{equation}
\psi(s)=\sum_{n=N_{0}}^{N}\log\left(1-\nu+\nu\left\langle e^{sK_{n}^{N}}\right\rangle \right).\label{eq:cumstochasticgeneral}
\end{equation}
Knowing the statistical properties of the propagator gives access
directly to the statistical properties of the total number of mutants.
Before applying this concept to specific cases, let us compute the
first two cumulant coefficients:
\begin{eqnarray}
\kappa_{1} & = & \nu\sum_{n=N_{0}}^{N}\left\langle K_{n}^{N}\right\rangle \label{eq:cum1sto}\\
\kappa_{2} & = & \mbox{\ensuremath{\nu}}\sum_{n=N_{0}}^{N}\left\langle \left(K_{n}^{N}\right)^{2}\right\rangle -\nu^{2}\sum_{n=N_{0}}^{N}\left\langle K_{n}^{N}\right\rangle ^{2}\label{eq:cum2sto}
\end{eqnarray}
The mean $\kappa_{1}$ is what we already had in the deterministic
case, where $\left\langle K_{n}^{N}\right\rangle =N/n$. Let us express
the second moment of $K_{n}^{N}$ as a function of its mean and variance
$V_{n}^{N}$ 
\[
\left\langle \left(K_{n}^{N}\right)^{2}\right\rangle =V_{n}^{N}+\left\langle K_{n}^{N}\right\rangle ^{2}
\]
Then 
\[
\kappa_{2}=\nu(1-\nu)\sum_{n=N_{0}}^{N}\left\langle K_{n}^{N}\right\rangle ^{2}+\nu\sum_{n=N_{0}}^{N}V_{n}^{N}
\]
The first term on the RHS of the above relation is what we already
had in the case of deterministic growth. The second term is the contribution
of the stochasticity of the propagator to the variance of the number
of mutants at population size $N$.

\subsection{Linear birth process.\label{sub:Linear-birth-process.}}

Consider the case where the growth of the WT is deterministic and
continuous
\[
\frac{dn}{dt}=\alpha(n,t)n
\]
while the mutant, once it has appeared, follows a stochastic growth
with transition probability density 
\[
\tilde{W}(m\rightarrow m+1)=\alpha(n,t)m
\]
where $\tilde{W}(m\rightarrow m+1)dt$ is the probability that this
lineage of the mutant has increased its size by one unit in the interval
$[t,t+dt]$. This model was first introduced by Lea and Coulson\cite{Lea1949}
for exponential growth case $\alpha(n,t)=\alpha$. 

A note of caution should be made here. Although widely used, the linear
birth model may not be very realistic, as bacterial division times
are not exponentially distributed. Indeed, after a division, a bacterium
needs to elongate again to its original size before being able to
divide again, so that the next division time cannot be smaller than
a finite time $\tau$. In fact, the distribution of division times
around the time $\tau$ is fairly narrow and the division process
is much less random than a linear birth process. The phenomenon has
been experimentally investigated by a microfluidic device by Wang
et al \cite{Wang2010} ; the overestimation of the mutation rate by
a linear birth model, for the exponential growth case, has been investigated
by Ycart \cite{Ycart2013}. 

Let us now come back to the linear birth model. As in the previous
section, the real time is not the best choice of independent variable
and we can write the stochastic growth of the mutant in terms of WT
population size: noting $W(m\rightarrow m+1)dn$ the probability that
a mutant has divided when the WT population size $\in[n,n+dn]$, we
have 
\begin{equation}
W(m\rightarrow m+1)=\tilde{W}(m\rightarrow m+1)\frac{dt}{dn}=\frac{m}{n}\label{eq:transition-n}
\end{equation}
Note that the equation for the mean of this lineage is 
\[
\frac{d\left\langle m\right\rangle }{dn}=\frac{\left\langle m\right\rangle }{n}
\]
which conserves the ratio between the size of this lineage and the
WT population, in agreement with the deterministic case investigated
above (subsection \ref{sub:Equal-growth-of}).

The master equation governing the growth of the mutant is 
\begin{eqnarray}
\frac{\partial P(m;n)}{\partial n} & = & W(m-1\rightarrow m)P(m-1;n)-\nonumber \\
 &  & W(m\rightarrow m+1)P(m;n)\label{eq:MasterEquation}
\end{eqnarray}
and the mgf of the propagator $K_{n}^{N}$ is given by (see appendix
\ref{sub:charKnN}): 
\begin{equation}
\left\langle e^{sK_{n}^{N}}\right\rangle =\frac{ne^{s}}{(n-N)e^{s}+N}\label{eq:charKnN}
\end{equation}
Using now the expression (\ref{eq:cumstochasticgeneral}) and the
continuous variable $x=n/N$, we obtain the cumulant generating function
of the number of mutants 
\begin{equation}
\psi(s)=N\int_{x_{0}}^{1}\log\left(1-\nu+\frac{\nu e^{s}x}{e^{s}(x-1)+1}\right)dx\label{eq:psistochastic}
\end{equation}
Note that the above integral can be expressed in an analytical, albeit
cumbersome, form. The expressions for the two first cumulant coefficients
are 
\begin{eqnarray*}
\kappa_{1} & = & -\theta\log x_{0}\\
\kappa_{2} & = & \frac{\theta\phi}{x_{0}}(2-\nu)+\theta\log x_{0}\\
 & \approx & 2\frac{\theta\phi}{x_{0}}+\theta\log x_{0}
\end{eqnarray*}
Where as before, $\theta=N\nu$ and $\phi=1-x_{0}$. The variance
of the number of the mutants is now approximately twice what we had
for the deterministic case. The exponential growth case can be recovered
from the above expression and is equal to the expressions given by
Lea and Coulson and Zheng (\cite{Zheng1999}, eq.52-53). 

Other cumulant coefficients can be readily recovered by multiple derivation
of expression (\ref{eq:psistochastic}). Restricting the computations
to the leading order of $\nu$ and $x_{0}$, the expression for the
cumulant coefficients is (see appendix \ref{sub:CumCoefSto}):
\begin{equation}
\kappa_{p}=\theta\frac{p!}{(p-1)x_{0}^{p-1}}\,\,\,\,\,p\ge2\label{eq:kpSto}
\end{equation}

Note that even in the case of exponential growth, no general expression
for the cumulant coefficients could be obtained by classical methods
(\cite{Zheng1999}). Comparing the above expression with the deterministic
case where $\kappa_{p}=N\nu/(p-1)x_{0}^{p-1}$, we see that indeed
the linear birth process induces large amplification of the $p-$th
cumulant coefficient by a factor of $p!$.

\paragraph{The probabilities.}

To compute the probabilities, it is more advantageous to use the probability
generating function (pgf)
\[
G(z)=\left\langle z^{X}\right\rangle =e^{\psi(z)}
\]
where $\psi(z)$ is defined from (\ref{eq:psistochastic}) by setting
$z=e^{s}$, \emph{i.e.}:
\begin{equation}
\psi(z)=N\int_{x_{0}}^{1}\log\left(1-\nu+\frac{\nu zx}{z(x-1)+1}\right)dx\label{eq:psiz}
\end{equation}
Approximating the above expression to the first order in $\nu$, we
have
\begin{equation}
\psi(z)=\theta(\frac{1}{z}-1)\log\left(1-\phi z\right)+O(\nu^{2})\label{eq:psiapproxZ}
\end{equation}
where $\theta=N\nu$ and $\phi=1-x_{0}$. We therefore obtain a simple
expression for the probability generating function 
\begin{equation}
G(z)=(1-\phi z)^{\theta(1/z-1)}\label{eq:Gz}
\end{equation}
For the case of exponential growth, the above expression (without
the $\phi$ factor) was first discovered by Lea and Coulson\cite{Lea1949};
omitting the $\phi$ factor however results in divergent moments.
The correct expression can be seen in the Zheng review (\cite{Zheng1999},
eq 65). We stress again that  relation (\ref{eq:Gz}) is very general
and does not depend on the assumption of exponential growth. Note
that in the case of exponential growth, $\theta\sim\exp(\beta t)$
and all the moments diverge as $t\rightarrow\infty$. This divergence,
discussed by Bartlett and later by Zheng \cite{Zheng1999}, cannot
be cured within the framework of the exponential growth model. No
such divergence exists in the present formulation, as the WT population
size, following any realistic growth curve, will remain finite. 

In order to evaluate the probabilities, we have first to compute $\psi^{(p)}(0)$.
Expanding expression (\ref{eq:psiapproxZ}) in powers of $z$ we have 

\[
\psi(z)=-\theta\phi+\theta\sum_{n=1}^{\infty}\left(\frac{\phi^{n}}{n}-\frac{\phi^{n+1}}{n+1}\right)z^{n}
\]
Keeping only the first leading terms in $\theta$ and $x_{0}$ we
have 
\begin{eqnarray}
\psi(z=0) & = & -\theta\phi\label{eq:psi0}\\
\frac{1}{p!}\psi^{(p)}(z=0) & = & \frac{\theta}{p(p+1)}-\frac{\theta px_{0}^{2}}{p+1}\label{eq:psip}
\end{eqnarray}
where the second term in $x_{0}^{2}$ can be neglected for $p\ll1/x_{0}$
and $\theta\phi\approx\theta$. We can now use the Faà Di Bruno Formula
\cite{Olver2010} to compute the $k-$th derivative of $G(z)$ at
$z=0$ and obtain the probabilities :
\begin{eqnarray*}
P(k) & = & \frac{1}{k!}G^{(k)}(0)\\
 & =e^{-\theta} & \sum_{\{m_{j}\}}\prod_{j=1}^{k}\frac{1}{\left(m_{j}\right)!}\left(\frac{\theta}{j(j+1)}\right)^{m_{j}}
\end{eqnarray*}
where the sum is taken over all $k-$tuples $\{m_{j}\}$ such that
$\sum_{j=1}^{k}jm_{j}=k$. The above formula can be easily programmed
to compute numerically the probabilities. The only linear term in
$\theta$ in the above formula is for $m_{k}=1$, $m_{i<k}=0$. Therefore,
for $\theta\ll1$, the probabilities take the simple form of 
\begin{eqnarray*}
P(0) & = & e^{-\theta}\\
P(k) & = & e^{-\theta}\frac{\theta}{k(k+1)}\,\,\,\,\,\,\,k>0
\end{eqnarray*}
The explicit expression for the probabilities in the linear birth
model has been intensely investigated by many authors, and reviewed
by Zheng (\cite{Zheng1999}, 5.3) and the above expressions are known
for the constant linear birth model. An explicit expression in terms
of Landau distribution has been found recently by Kessler and Levine
\cite{Kessler2013,Kessler2014} for both the deterministic and stochastic
growth model. 

Experimentally, obtaining the probabilities implies a very large number
of parallel cultures and the above expressions may not be of great
practical use.

\paragraph{Different growth rate.}

The discussion of the preceding subsection can be extended to take
into account a different relative growth rate for the mutant compared
to the WT: 
\begin{eqnarray*}
\frac{dn}{dt} & = & \alpha(n,t)n\\
\tilde{W}(m\rightarrow m+1) & = & c\alpha(n,t)m
\end{eqnarray*}
Repeating the discussion of the preceding sections, the cumulant generating
function in this case is (see appendix \ref{sub:charKnN}) 
\begin{equation}
\psi(s)=N\int_{x_{0}}^{1}\log\left(1-\nu+\nu\frac{x^{c}e^{s}}{(x^{c}-1)e^{s}+1}\right)\label{eq:cgfc}
\end{equation}
from which the cumulant coefficients can be deduced as before.

\subsection{random relative growth rate.\label{sub:Variable-growth-rate.}}

The traditional formulation of LD distribution assumes that all mutants
are similar in their growth function. As many different mutations
can bring a bacteria to the same phage resistance, this assumption
may seem too restrictive and can be easily relaxed. For example, a
comprehensive study of this phenomenon has recently been published
\cite{Stiffler2015} where the growth rate of \emph{all} mutants in
the gene TEM-1, conferring resistance to the antibiotic cefotaxime,
where measured and shown to be variable in some conditions. 

Consider the case where the growth of both mutants and WT are deterministic
as in subsection \ref{sub:Different-growth}
\begin{figure}
\begin{centering}
\includegraphics[width=0.85\columnwidth]{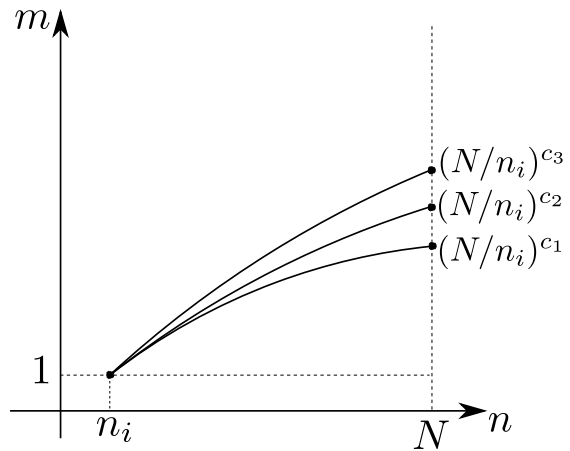}
\par\end{centering}

\protect\caption{Deterministic growth, random relative growth rate. The size of the
lineage of mutant $i$ appearing at WT population size $n_{i}$ is
$m_{i}(n)=(n/n_{i})^{c}$ for $n\ge n_{i}$. This time however, $c$
is a random variable. \label{fig:Deterministic-growth,-random}}

\end{figure}
\[
\frac{dn}{dt}=\alpha(n,t)n\,\,;\,\,\frac{dm}{dt}=c\alpha(n,t)m
\]
but the relative growth rate $c$ is now a random variable (figure
\ref{fig:Deterministic-growth,-random}): when it appears, a mutant
picks a relative growth rate $c$ from a given distribution, which
is transmitted to its progeny. 

Following the discussion of subsection \ref{sub:General-discussion.},
the propagator this time is 
\[
K_{n}^{N}=(N/n)^{c}
\]
where $c$ is now a random variable. Let us denote $\rho(s)$ its
moment generating function 
\[
\rho(s)=\left\langle e^{sc}\right\rangle 
\]
Then according to relations (\ref{eq:cum1sto}-\ref{eq:cum2sto}),
the first two cumulants are now, to the first order in $\nu$:
\begin{eqnarray*}
\kappa_{1} & = & \theta\int_{0}^{-\log x_{0}}\rho(z)e^{-z}dz\\
\kappa_{2} & = & \theta\int_{0}^{-\log x_{0}}\rho(2z)e^{-z}dz
\end{eqnarray*}

For example, if $c$ follows a Normal distribution $c={\cal N}(\mu,\sigma)$,
then $\rho(s)=\exp(\mu s+(1/2)\sigma^{2}s^{2})$ and these expressions
can be evaluated in terms of the error function. For the specific
case of $\mu=1$ and $\sigma\ll1$, restricting the computation to
the leading orders of $\nu$ and $x_{0}$, we have 
\begin{eqnarray*}
\kappa_{1} & = & -\theta\log x_{0}\left(1+\frac{\sigma^{2}}{6}(\log x_{0})^{2}\right)\\
\kappa_{2} & = & \frac{\theta}{x_{0}}\left(1+2\sigma^{2}\left(1+(1+\log x_{0})^{2}\right)\right)
\end{eqnarray*}

\section{Discussion and Conclusion.}

In the previous sections we have given the general solution of the
LD distribution through the derivation of its cumulant generating
function. The key point to this derivation is to change the independent
variable from time to WT population size, which is indeed the relevant
variable : whatever the growth function, a mutation can occur only
when a WT cell divides and WT population size changes from $n$ to
$n+1$. This consideration considerably simplifies the solution of
the problem and allows us to extend the solution to arbitrary growth
curves for the WT population. The mathematical formulation applies
equally well to the case of deterministic and stochastic growth. 

This mathematical formulation is sufficiently simple to allow for
many generalizations, some of which have been considered in this article:
for example variable mutation probability (subsection \ref{sub:Variable-mutation-probability.})
or random relative growth rate (subsection \ref{sub:Variable-growth-rate.})
have been investigated.

Other generalizations that we have not developed can be considered.
For example, for the stochastic growth case, we have only considered
the linear birth process. Other, more realistic cases can be envisaged
where the distribution of the division times are not exponential,
along the lines developed by Ycart\cite{Ycart2013}. One can also
consider the case where both mutants and WT grow stochastically. These
generalizations would be straightforward if the moment generating
function of the propagator $K_{n}^{N}$ can be derived in explicit
form. Another possible generalization would be to take into account
the experimental uncertainty on the initial and final value of the
WT population, and its influence on the estimation of mutation probability,
as has been considered by Ycart and Veziris\cite{Ycart2014a}. 

To summarize, we have developed in this article a versatile method
for investigating the Luria Delbrück distribution with an arbitrary
growth function. The method uses only very few measurable parameters,
namely the initial and final number of the WT population. We believe
that the method we propose here can be used as a simple basis for
further investigations of the LD distribution.

\appendix*

\section{Mathematical details.}

\subsection{Geometric distribution of mutants appearing at WT population size
$n$.\label{sub:Geometric-distribution-of}}

We have noted $\nu$ the probability of apparition of a mutant when
\emph{a} WT cell divides, and $X_{n}$ the random variable tracking
the number of mutants appearing when WT population changes its size
from $n$ to $n+1$. The probability of producing $k$ mutant during
this change is geometrically distributed:
\[
\mbox{Pr}(X_{n}=k)=(1-\nu)\nu^{k}
\]
As $\nu\ll1$, we have approximated $X_{n}$ by a binary process ($X_{n}=0,1$)
in the article. This constraint can be relaxed. Following notations
of subsection \ref{sub:Equal-growth-of}, 
\[
\left\langle e^{sY_{n}^{N}}\right\rangle =\frac{1-\nu}{1-\nu e^{sN/n}}
\]
 and 
\begin{eqnarray*}
\psi(s) & = & \sum_{n=N_{0}}^{N}\log\left(\left\langle e^{sY_{n}^{N}}\right\rangle \right)\\
 & = & N\int_{x_{0}}^{1}\log\left(\frac{1-\nu}{1-\nu e^{s/x}}\right)dx\\
 & = & -\theta\phi-N\int_{x_{0}}^{1}\log\left(1-\nu e^{s/x}\right)dx
\end{eqnarray*}
where $\theta=-N\log(1-\nu)$ and $\phi=1-x_{0}$. The above expression
is equal to expression (\ref{eq:cgf-deterministic}) to the first
order in $\nu$.

\subsection{Moment generating function of $Y_{n}^{N}$ in the stochastic case.\label{sub:appa}}

Consider the random variable $Y_{n}^{N}=X_{n}K_{n}^{N}$ where $X_{n}$
is a boolean variable $P(X_{n}=0)=1-\nu$ and $P(X_{n}=1)=\nu$ (subsection
\ref{sub:General-discussion.}); For simplicity, we suppose that $K_{n}^{N}$
is a positive discrete random variable. Then, 
\begin{eqnarray*}
P(Y_{n}^{N}=0) & = & 1-\nu+\nu P(K_{n}^{N}=0)\\
P(Y_{n}^{N}=k\ne0) & = & \nu P(K_{n}^{N}=k)
\end{eqnarray*}
The mgf is then 
\begin{eqnarray*}
\left\langle e^{sY_{n}}\right\rangle  & = & \sum_{k=0}^{\infty}P(Y_{n}=k)e^{sk}\\
 & = & 1-\nu+\nu P(K_{n}^{N}=0)+\\
 &  & \sum_{k=1}^{\infty}\nu P(K_{n}^{N}=k)e^{sk}\\
 & = & 1-\nu+\nu\left\langle e^{sK_{n}^{N}}\right\rangle 
\end{eqnarray*}
which is the expression (\ref{eq:charYn}).

\subsection{Moment generating function of the propagator $K_{n}^{N}$.\label{sub:charKnN}}

A mutant appearing in one copy when the WT population size is $n_{0}$
will reach size $K_{n_{0}}^{n}$ when the WT population reaches size
$n$. The forward master equation governing the probabilities of the
propagator, $\mbox{Pr}(K_{n_{0}}^{n}=m)$ is given by equation (\ref{eq:MasterEquation}).
The mgf of the propagator, $\phi(s,n)$ therefore obeys the following
equation :
\begin{equation}
n\frac{\partial\phi}{\partial n}+(1-e^{s})\frac{\partial\phi}{\partial s}=0\label{eq:pgfeq}
\end{equation}
with the boundary conditions 
\begin{eqnarray*}
\phi(s,n=n_{0}) & = & e^{s}\\
\phi(s=0,n) & = & 1
\end{eqnarray*}
Equation (\ref{eq:pgfeq}) is a linear first order partial differential
equation that can be solved by the method of characteristics:
\[
\phi(s,n)=\frac{(n_{0}/n)e^{s}}{(n_{0}/n-1)e^{s}+1}
\]
Changing now the notation to denote $n$ as the WT population size
of the mutant occurrence, and $N$ the final size of the WT population,
we obtain the expression (\ref{eq:charKnN}).

If the relative growth rate of the mutant is not 1 but $c$, where
$c$ is an arbitrary constant, the transition probability for the
mutant once it has appeared is 
\[
W(m\rightarrow m+1)=c\frac{m}{n}
\]
and $\phi(s,n)$ obeys the following equation: 
\begin{equation}
\frac{1}{c}n\frac{\partial\phi}{\partial n}+(1-e^{s})\frac{\partial\phi}{\partial s}=0\label{eq:phic}
\end{equation}
This equation can be transformed into equation (\ref{eq:pgfeq}) by
a simple scaling $n\rightarrow n^{c}$ and the mgf is therefore given
by 
\[
\phi(s,n)=\frac{(n_{0}/n)^{c}e^{s}}{\left((n_{0}/n)^{c}-1\right)e^{s}+1}
\]
from which the cumulant generating function of the number of mutants
can be deduced.

\subsection{Cumulant coefficients for the stochastic growth.\label{sub:CumCoefSto}}

To the first order in $\nu$, the cgf for the linear birth model (subsection
\ref{sub:Linear-birth-process.}) is given by 
\[
\frac{1}{\theta}\psi(s)=(e^{-s}-1)\log\left(1-e^{s}+x_{0}e^{s}\right)
\]
Expanding the above function into powers of $(1-e^{-s})$, we have
\begin{eqnarray*}
\psi(s) & = & (s+\log x_{0})(e^{-s}-1)+\\
 &  & \sum_{n=2}^{\infty}\frac{1}{n-1}\frac{(1-e^{-s})^{n}}{x_{0}^{n-1}}
\end{eqnarray*}
Evaluating the $p-th$ derivative at $s=0$, we have 
\[
\psi^{p}(0)=\frac{p!}{(p-1)x_{0}^{p-1}}+O(\frac{1}{x_{0}^{p-2}})
\]
 which is the expression given in equation (\ref{eq:kpSto}).

\paragraph{Acknowledgements.}

I sincerely thank Marcel Vallade, Olivier Rivoire and Erik Geissler
for the critical reading of the manuscript.

\bibliographystyle{unsrt}

\end{document}